\documentclass{aa}
\usepackage{epsfig}
\def\beq{\begin{equation}}
\def\eeq{\end{equation}}
\begin{document}

\title{Waves and instabilities in an anisotropic universe}

\author{Demetrios Papadopoulos \inst{1},
  Loukas Vlahos \inst{1}
  \and
  F. Paul Esposito \inst{2}}
\institute{Department of Physics, Aristoteleion University of
Thessaloniki, \\54006 Thessaloniki, Greece
\and Department of Physics, University of Cincinati, Cincinaty, OH 45219
}

\offprints{D. Papadopoulos}

\date{Received .........; accepted .............}
\titlerunning{Waves in an anisotropic universe}
\authorrunning{D. Papadopoulos et al....}

\abstract {The excitation of low frequency plasma waves in an
expanding anisotropic cosmological model which contains a magnetic
field frozen into the matter and pointing in the longitudinal
direction is discussed.  Using the exact equations governing finite-amplitude
wave propagation in hydromagnetic media within the framework of
general theory of relativity, we show that a spectrum of magnetized sound waves
will be excited and form large scale ``\textit{damped oscillations}'' on the expanding universe.
The characteristic frequency of the excited waves is slightly shifted away from
the sound frequency and the shift depends on the strength of the primordial
magnetic field. This magnetic field dependent shift
may  have an effect on the acoustic peaks of the CMB.
}

\maketitle

\section{Introduction}
At present not much is known about the existence or absence of
magnetic fields in the early Universe. Several scenarios exist for
the generation of magnetic fields (see  review by Grasso \& Rubinstein 2001). The
implications of the primordial magnetic fields for the formation
and evolution of the observed structures have been the subject of
continuous theoretical investigation.

The observed magnetic field in the clusters and galaxies is in
energy equipartition with the gas and the cosmic rays in these
systems. The observed magnetic field may be primordial in nature,
being left over from the early universe, or created when the first
structures formed. Even if we assume the presence of a
protogalactic dynamo to explain the magnitude of the magnetic
field still requires a small primordial field. It is a difficult
task to determine if the origin of the cluster or the galactic
magnetic field is primordial. In contrast the role of the magnetic
field in the fluctuations spectrum of the cosmic microwave
background anisotropies can provide bounds in the large scale
magnetic field of the early universe (see Adams et al. 1996,
Barrow et al. 1997, Enqvist 1997, Olinto 1997, Durrer et al. 1998).

Many recent studies used a Newtonian or a
Friedmann-Robertson-Walker (FRW) model for the evolving Universe
and super-impose a large scale ordered magnetic field. The
magnetic field is assumed too weak to destroy the Robertson-Walker
isotropy. The anisotropy induced by the magnetic field is treated
as perturbation (Durrer et al. (1998), Ruzmakina \& Ruzmakin 1971, Tsagas \& Barrow 1998).
Current observations give a strong motivation for the adoption of
a FRW model (see reviews above) but the uncertainties on the
cosmological standard model are several and the limits of the
approximations and the effects one losses by neglecting the
anisotropy of the background magnetic field should be
investigated.

In this article, departing from the traditional approach, we adopt
an anisotropic cosmological model, developed initially by
Thorn(1967) and Jacobs(1968). In this model the magnetic field is responsible for
the initial anisotropic expansion of the universe. The model is
homogeneous and has two equivalent ``transverse" directions and
one inequivalent ``longitudinal" direction at each point in space
time. It contains a perfect fluid obeying a ``stiff''  equation of
state $p=\gamma \rho,$ with $1/3\leq \gamma\leq 1.$ The magnetic
field is frozen into the matter and directed along the
longitudinal direction.

We analyze the stability of the linear perturbations using the
above model. Magnetized density perturbations were studied in detail in \cite{Kim}
using a Newtonian model for the evolving Universe.  Several
authors used the FRW model to study magnetized perturbations in
the radiation era (see Holcomb \& Tajima 1989),
 the dissipation of MHD waves in
recombination era  (Jedamzik et al.2000) and  in the inflanatory era
(Brevik \& Sandvik 2000). The propagation of the magnetosonic waves in
sub-horizon and super-horizon scales were discussed by Tsagas \& Maartens (2000a) and more
recently a magnetized Biannchi I background model was used to study the
coupling of the magnetism with the geometry (Tsagas \& Maartens 200b).

The main scope of this study is to search for the spectrum of the
unstable low frequency plasma waves in an anisotropic model for
the universe, using the formalism developed by Papadopoulos \& Esposito (1982). In
section 2 we present the basic equations used and in section 3 we
perturb the MHD equations keeping first order terms and searching
for solutions when the perturbed quantities have the form of a
plane wave. We derived the linear dispersion relation for the
magnetosonic waves and search for the spectrum of excited waves.

\section{Basic equations}
The general relativistic version of the magnetohydrodynamic
equations are:
 \begin{eqnarray} \label{Bas1}
 \dot{x} u^a&+&x\dot{u}^a+x\theta
u^a+(p+\frac{H^2}{2})_{;b}g^{ab}\nonumber\\&-&(H^a
H^b)_{;b}=0,\end{eqnarray}
\begin{eqnarray}\label{Bas2}
& &(\epsilon-\frac{H^2}{2})_{;ab} u^a
u^b=h^{ab}(p+\frac{H^2}{2})_{;ab}+2\dot{(H^2\theta)}\nonumber\\&-&
(H^aH^b)_{;ab}
+2x(\frac{2\theta^2}{3}+\sigma^2-\omega^2-\dot{u}^a\dot{u}_{a})\nonumber
\\&+&\frac{x}{2}(\rho+3p+H^2)+2\dot{u}_{a}
(H^aH^b)_{;b}\nonumber\\&+&(H^2)_{;a}\dot{u}^a,  \end{eqnarray}
 and
 \beq \label{Bas3}
 \dot{\frac{\mu
H^2}{8\pi}}=\frac{\mu}{4\pi}\sigma_{ij} H_i
H_j-\frac{4\theta}{3}(\frac{\mu H^2}{8\pi}), \eeq where c=G=1,$\mu$ is
the permeability, $u^a$ is the fluid velocity, $\rho$ is the mass
density, $\epsilon=\rho+\rho\Pi$, $\rho\Pi$ is the internal
energy,
  $\dot{u}^a=u_{;c}^a u^c$, $\theta=u^a_{;a}$ is the expansion velocity,
   $x=\epsilon+p+H^2$, H the magnetic field
and  $h^{ab}=g^{ab}+u^a u^b$. In \cite{Pap} the Eqs.
(\ref{Bas1}), (\ref{Bas2}), and (\ref{Bas3}) were perturbed using
also the the condition $\delta g_{ab}=0$ known as Cowling approach
\cite{Cow}, and keeping only first order terms in pressure,
density, velocity, and magnetic field (see Eqs (47), (48) and (49)
in Papadopoulos \& Esposito 1982).

We linearize the perturbed and search for the amplification or
damping of small amplitude hydromagnetic waves in the early
universe described by the well known anisotropic cosmological
model (Thorn 1967), \beq \label{met} ds^2=-dt^2+A^2
(dx^2+dy^2)+W^2 dz^2; \eeq here, (i) the model contains perfect
fluid obeying the equation of state $p=\gamma \rho$ where
\mbox{$\frac{1}{3}<\gamma\leq 1$}, (ii) the fluid is assumed to be
co-moving  with the coordinate system and therefore
$u^{\mu}=(-1,0,0,0)$ and (iii) as seen in the rest frame of the
fluid there is a magnetic field of strength $H$ pointing in the
$z$-direction but no electric field. The functions A and W which
enter into line element are $A=A(t)=t^{1/2}$ and $W=W(t)=t^l$,
while
\[ \rho=\frac{3-\gamma}{16\pi t^2 (1+\gamma)},\]
\[
H=\frac{(1-\gamma)^{1/2} (3 \gamma-1)^{1/2}}{2 t (1+\gamma)},
\]
and $l=(1-\gamma)/(1+\gamma). $

The ratio of the cyclotron ($\omega_b$) to plasma ($\omega_p$)
frequency is a function of $\gamma$ $(\omega_b^2/\omega_p^2)=
(2(1-\gamma)(3\gamma-1))/(3-\gamma). $ The
cyclotron frequency tends to zero as $\gamma \rightarrow 1$ or
$\gamma \rightarrow 1/3$ and the ratio $0\leq
\omega_b^2/\omega_p^2\leq 0.3 $ remains bounded when $1/3 \leq
\gamma \leq 1.$

The Standard Hot Big Bang (SHBB) is roughly divided into two
regimes, the radiatin dominated era, for which the $\gamma=1/3,$
and the (pressurless) matter dominated era, for which $\gamma
\simeq 0.$ Thus, at first, it seems not possible for the Universe
to experience a stiffer equation of state, with $1/3< \gamma\leq
1$. However, is simple extension of SHBB it is possible to
acchieve this in a rather natural way.

One prominent example is cosmologies which include a scalar field
component in the matter content of the Universe. A homogeneous
scalar field $\phi=\phi(t)$ can be treated as an ideal gas with
density $\rho_\phi=\rho_{kin}+V$ and pressure
$p_\phi=\rho_{kin}-V$, where $\rho_{kin}=(1/2) \dot{\phi}^2$ and
$V=V(\phi)$ are the field's kinetic and scalar potential energy
respectively. In the case $V>>\rho_{kin},$ then $\gamma_\phi\simeq
-1$ and, if $\rho_\phi$ dominates the Universe then the latter is
forced to undergo accelerated expansion. Such a situation, if it
taken to occur at the very first stages of the Universe evolution,
is the typical realization of inflation, in which $\phi$ is called
the inflaton field (Olive 1990). At the end of the inflationary
period $\phi$ usually oscillates around the minimum of $V(\phi)$
and decays into other particles creating the thermal bath of the
SHBB. However, there exist non-oscillatory models of inflation in
which the inflation does not decay at the end of the inflation
but, instead, it engages into a rapid roll-down of its steep
potential (Felder et al. 1999). In such models during this phase the
Universe continues to be dominated by the scalar field but this
time we have $V<< \rho_{kin}$ so that $\gamma_\phi\simeq 1$.This
period is usually refereed to as kination or deflation
(Joyce \& Procopec 1998). Eventually kination ends and the SHBB begins. At the
transition between kination and radiation domination we have a
change of $\gamma$ from $1 \rightarrow 1/3$.

In the next section we derive the dispersion relation for the low
frequency waves, assuming that all the perturbed quantities
$(\delta \rho, \delta p, \delta H^\mu,\delta u^\mu)$ can be
expressed in the form $e^{i(nt-kz)}.$

\section{Excitation of low frequency plasma waves}

We assume that the wave vector and the external magnetic field are
along the z-axis $ H^{\mu}=(0,0,H^3)$ , $k^{\mu}=(0,0,0,k)$, and
the perturbations of the fluid velocity and the magnetic field
have a general form $\delta H^{\mu}=(\delta H^1, \delta H^2,\delta
H^3), ~~ \delta u^{\mu}=(\delta u^0, \delta u^1, \delta u^2,
\delta u^3)$. After some long but
straight forward calculations, we derive the generalized
dispersion relation,
\begin{eqnarray} \label{Dis1}
-n^2&+&k^2c_s^2-(1+c_s^2)[-\frac{8}{3}\theta^2+2\sigma^2+\frac{1}{2}(\rho+3p)\nonumber\\
&+&\frac{2H_{,0}^2\theta}{\rho+p}+\frac{4\theta^2
H^2}{3(\rho+p)}-
\frac{2H^2}{3}]\nonumber\\&-&\frac{1}{2}(p+\rho)(1+3c_s^2)\nonumber\\
&-&in[4\theta+\frac{2H_{,0}^2}{(\rho+p)}-
\frac{4\theta H^2}{3(\rho+p)}]\nonumber\\
&=&(\mathcal{R}_1+in \mathcal{R}_2)\frac{\delta
H^3}{\mathcal{\delta} \epsilon},
\end{eqnarray}
and
\begin{eqnarray}\label{frc1}
\frac{\delta H^3}{\delta
\epsilon}=\frac{H^3}{4\pi(2-\gamma)}[1-\Lambda_1-in\Lambda_2],
\end{eqnarray}
 where the functions $\mathcal{R}_1,\mathcal{R}_2,
\Lambda_1, \Lambda_2$ and $\mathcal{D}$ are:\\
\begin{eqnarray*}\mathcal{R}_1&=&[H_{3,00}-4(\theta H_{3,0}+\theta_{,0}H_3)-2H_{,0}^3\Gamma_{33}^0]
\\&+&H_3[-\frac{4}{3}H^2+\frac{4}{3}\theta^2+2\sigma^2+2(\rho+2p)]\\&+&g^{33}[-2\Gamma_{33,0}^0
-2\Gamma_{33}^0\Gamma_{b0}^b+2\Gamma_{33}^0\Gamma_{30}^3]H_3\\&-&(n^2+k^2)H_3
,\end{eqnarray*}
\[\mathcal{R}_2=2H_{3,0}-4\theta H_3-\Gamma_{33}^0 H_3
g^{33},\]
\begin{eqnarray*}\Lambda_1&=&\frac{1}{\mathcal{D}}[k^2(H^3)^2+(2\gamma-1)\theta[2x\Gamma_{30}^3+
(p+H^2/2)_{,0}]\\&-&k^2c_s^2g^{33}(\epsilon+p)],\end{eqnarray*}
\[\Lambda_2=(2\gamma-1)x\theta,\]
\begin{eqnarray*}
\mathcal{D}&=&-n^2
x+k^2(H^3)^2-(2-\gamma)\theta\left[2x\Gamma_{30}^3+(p+\frac{H^2}{2})_{,0}\right]
\\&-&in\left [2x\Gamma_{30}^3+p+\frac{H^2}{2}-(2-\gamma)x\theta\right].\end{eqnarray*}

In the absence of magnetic field (e.g $\gamma=1/3$) the model used here
reduces to a the standard FRW model without cosmological constant, since $A(t)=W(t)=t^{1/2}$ and the real part of the
dispersion relation (Eq. (\ref{Dis1})) has a very simple form
\begin{eqnarray}\label{Dis4}
n^2=k^2c_s^2&+&(1+c_s^2)[\frac{8}{3}\theta^2-\frac{1}{2}(\rho+3p)]
\nonumber\\&-&\frac{1}{2}(p+\rho)(1+3c_s^2).
\end{eqnarray}
Eq.(\ref{Dis4}) reduces further if we use an Enstein cosmalogical model
with cosmological constant  $\Lambda=(1/2)(\rho+3p),$
which makes the medium static and homogeneous
and $\theta=0.$ Using the above assumptions
the Jeans instability $(n^2<0$)
for a static isotropic and homogeneous model was derived by in Jackson (1972).
The Jeans instability is quenched by expansion ($\theta \neq 0$) in the FRW model, without
cosmological constant,
since
the second
term in the right hand side of Eq. (\ref{Dis4}) is always positive and dominates
over the third which drives the Jeans instability in the static model.

 It is easy to verify
that  the dispersion relation (Eq. \ref{Dis1})
is simplified considerably in the limit $t\rightarrow 0,$
\begin{eqnarray} \label{Dis2}
&-&n^2[1-\frac{u_A^2}{2-\gamma}]+k^2c_s^2+k^2u_{A}^2\frac{1}{2-\gamma}\nonumber\\&+&
T_1(\gamma,t)+inT_2(\gamma,t)=0,
\end{eqnarray}
where $c_s$ is the sound speed and $u_A^2={H^2}/{4\pi \rho}$ is the
Alfven speed. The functions $T_1$ and $T_2$ have the following functional form
\[ T_1(\gamma,t)=\frac{\tilde{T}_1(\gamma)}{t^2}\]
  \begin{eqnarray*}\tilde{T}_1(\gamma)&=&-(9\gamma^6-993\gamma^5+4842\gamma^4
-6818\gamma^3+2889\gamma^2\\&-&461\gamma+788)/(6(\gamma-2)(\gamma-3)^2(1+\gamma)^2)
 \end{eqnarray*}
 and
\begin{eqnarray} \label{T2}
T_2(\gamma,t)&=&-4\theta+\frac{4\theta}{3}\left(\frac{H^2}{4\pi(\rho+p)}\right)+
4\theta\left(\frac{H^2}{4\pi(2-\gamma)}\right)\nonumber\\&-&\frac{2H_{,0}^2}{4\pi(\rho+p)}
-\frac{2H_{3,0}H_3}{4\pi\rho(2-\gamma)}\nonumber\\&+&
\frac{(1-\gamma)}{(\gamma+1)t} \left(\frac{H^2}{4\pi\rho(2-\gamma)}\right).
\end{eqnarray}
where $\theta=-2/((1+\gamma)t)$.The function $T_2$ depends on $\gamma$ and t since
the magnetic field, the pressure and density
are all functions of $\gamma$ and t. In other words $T_2(\gamma,t)$ has the form
\[ T_2(\gamma,t)=\frac{C(\gamma)}{t},\] where
\[C(\gamma)=
\frac{45\gamma^4-36\gamma^3-126\gamma^2+
212\gamma-127}{3(\gamma-3)(\gamma-2)(\gamma+1)^2}.\]
We can easily show that $T_1(\gamma,t)$ is positive and $T_2(\gamma,t)$
negative for all allowed values of
$1/3 \leq \gamma \leq 1.$

We re-write Eq. (\ref{Dis2}) in the form $D_r+iD_i=0$ and assuming
that the the real part of the excited frequency is much larger
than the imaginary part, $n_r>>n_i$, we obtain the frequency of
the excited wave from the equation $D_r=0,$
\begin{eqnarray} \label{Dist3}
n_{r}^2[1-\frac{u_A^2}{2-\gamma}]&=& k^2c_s^2+\frac{k^2u_{A}^2}
{2-\gamma}\nonumber\\&+&\frac{\tilde{T}_1(\gamma)}{t^2}.
\end{eqnarray}
and the imaginary frequency from the well known relation
$n_i=-D_{i}(n_r,k)/(\partial D_{r}(n_r,k)/\partial n_r)$(see
Krall \& Trivielpiece 1973)
\begin{eqnarray} \label{Im1}
n_i&=&\frac{T_2(\gamma,t)}{2(1-u_A^2(t)/(2-\gamma))}\nonumber\\&=&\left(\frac{C(\gamma)}{t}\right)
\left(\frac{1}{2(1-u_A^2(t)/(2-\gamma))}\right),
\end{eqnarray}
were $n_i$ has a large negative value when $ t \rightarrow 0$ and decay fast at later times.
The role of magnetic field is to make the expanding universe even more stable
against the expected Jeans type instabilities discussed early.

The real frequency on the other hand is shifted away from the sound speed
\begin{equation}\label{sound}
  n_r^2=\frac{k^2c_s^2+ \frac{k^2 u_A^2}{2-\gamma}+\frac{\tilde{T}_1}{t^2}}
  {1-\frac{u_A^2(t)}{2-\gamma}}
  \end{equation}
as $\gamma \rightarrow 1/3$ the magnetic field approaches
 zero and the anisotropic model used here approaches the  weakly magnetized FRW
 model used extensively in the literature. The characteristic
frequency $n_r$ will approach the frequency  $n_r=kc_s +\Delta n_r (u_A).$
The presence of the \`{A}lfven speed in the frequency
$n_r$ in Eq. (\ref{sound}) may be responsible for the
distortion of acoustic peaks, as it was pointed out first by Adams et al. (1996).

The imaginary part
is negative $n_i<0$ and a spectrum of low frequencies will be
excited with frequency $n_r.$ The linear \textbf{density
perturbation} will have the form
\beq
\delta\rho  \sim  \left[\rho_0 e^{|B(\gamma,t)|}\right]
e^{i(n_r t-k z)}\eeq
were $B(\gamma,t)= C(\gamma)/(2(1-u_A^2(t)/(2-\gamma))).$ $|B(\gamma,t)|$ is a decaying function of
time. For $t \rightarrow 0,$ $\gamma \rightarrow1$ then $u_A\rightarrow 1$
and $|B(\gamma,t)|  \rightarrow \infty$.
As $t>>0$ and $\gamma\rightarrow 1/3$ then $u_A \rightarrow 0$ and $|B(1/3,t)|\rightarrow (3/2).$
We then conclude that
in the early universe the amplitude becomes large and
 the excited waves
will form a \textbf{spectrum of damped oscillations} in the expanding
magnetized universe. We can search for the source of damping by analyzing the functional form of
 $T_2(\gamma,t)$ in Eq. (\ref{T2}). In the absence of magnetic field ($\gamma=1/3$)
 $T_2=-4\theta$ and the expanding plasma stabilizes the ion acoustic waves, on the other
 hand if the universe is static ($A=constant$ and $\theta\sim \dot{A}/A \sim0$)
 the stabilization of the magnetosonic waves is caused by the
 magnetic pressure.  In a
 magnetized expanding universe (Eq. (\ref{T2})) both processes
 will be combined to damp the  excited large amplitude oscillations.
Assuming, for the sake of argument that the universe at a certain time $t$=const
is becoming static,  a Jeans type instability sets in so we can easily realize that
 the damping is caused by the
 free energy available both in the plasma and magnetic pressure during the
 initial stages of its expansion.

\section{Summary}
We analyzed the wave propagation in the early Universe using an
anisotropic cosmological model which contains a magnetic field
frozen into the perfect fluid with equation of state $p=\gamma
\rho$ (with $1/3\leq \gamma \leq 1$) and pointing in the
z-direction. The choice of a stiff equation of state with $1/3<
\gamma \leq 1$ may be  realistic representation for the evolving
universe in the phase between the kination and the radiation era.

The anisotropic model for the early universe used can approach the weakly magnetized FRW model when
$\gamma \rightarrow 1/3$. In this case the anisotropic model used here approaches the isotropic
FRW model were the magnetic field is a small linear perturbation (Durrer et al.1998,
 Ruzmakina \& Ruzmakin 1971).
Our main results in this article are:
(1) The Jeans instability, found early for an Einstein universe
with very particular cosmological constant (Jackson 1972),
will be absent since the expansion and the presence of magnetic
field will act as a stabilizing force.
(2) The anisotropic magnetized Universe re-enforces the stability found initially in
the FRW model.
(3) A spectrum of low frequency and large amplitude  damped oscillations will appear in the early universe
with characteristic frequency $n_r=kc_s+\Delta n,$
where $\Delta n$ depends on the strength of the primordial magnetic field will be excited.
This magnetic field
dependent shift may cause measurable distortion of accoustic peaks (see also Adams et al.1996,
Jedamzik et al. 2000, Koh \& Lee 2000, Durrer et al. 2001).
The amplitude of the excited waves is large in the early universe and gradually decay.

We then conclude that
if the universe have passed through a strongly magnetized anisotropic phase, before the
recombination era started, the
waves predicted here will be responsible for the formation of large scale
fluctuations, as it has been shown in the 2-D numerical simulations (Brandenburg et al. 1996).

\begin{acknowledgements}
We grateful to our colleagues Konstantinos Dimopoulos and Christos
Tsagas for many stimulating discussions and for making several
comments which improved considerably our article.
\end{acknowledgements}

\end{document}